\documentclass[11pt]{cernrep}
\usepackage{epsfig}
\usepackage{graphicx}
\usepackage{cite,./mcite}
\begin{document}
\title{Heavy quark production at HERA and the LHC}
\author{Matthew Wing}
\institute{University College London and DESY}
\maketitle
\begin{abstract}
Measurements of heavy quark production, particularly from HERA, their theoretical 
understanding and their relevance for the LHC are reviewed\footnote{Since the 
presentation, some results have been updated; these are used in what follows.}. 
The status of beauty and charm production is discussed in the context of the 
different components of the production process: the parton density function of 
the colliding hadrons; the hard scatter; and the fragmentation of the quarks into 
hadrons. The theory of QCD at next-to-leading order generally describes well the 
hadronic structure and the production of heavy quarks although sometimes fails in 
details which are highlighted. The fragmentation of heavy quarks measured at HERA 
is consistent with that at LEP and hence supports the notion of universality.
\end{abstract}

%
%

\section{Why study heavy quark production?}

\begin{figure}[htp]
\begin{center}
~\epsfig{file=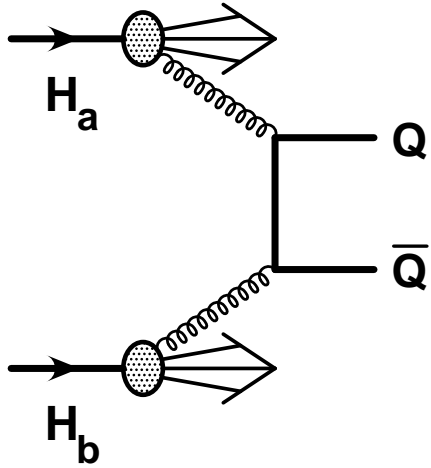,height=6cm}
\end{center}
\label{fig:generic}
\caption{Example of the production of a heavy quark pair in the collision of two hadrons.}
\end{figure}

The measurement of heavy quarks can give insights into many physical phenomena such 
as: new particles which are expected to decay predominantly to beauty (and charm); 
precise measurements of electroweak parameters; and, the subject of this paper, a 
deeper understanding of the strong force of nature. The strong force as described 
within perturbative Quantum Chromodynamics (QCD) should be able to give a precise 
description of heavy quark production. This postulate is described and tested here. 
The measurement of heavy quark production also yields valuable information on the 
structure of colliding hadrons. The production of a pair of heavy quarks in a generic 
hadron collision is shown in Fig.~\ref{fig:generic} where it can be seen 
that the process is directly sensitive to the gluon content of the hadron. Most 
information on the structure of a hadron comes from inclusive deep inelastic 
scattering where the gluon content is determined in the evolution of the QCD 
equations. Therefore measurement of such a process in Fig.~\ref{fig:generic} 
provides complimentary information to that from inclusive measurements. 

As well as understanding for its own sake, knowledge of the structure of hadrons 
will be important at future colliders such as the LHC and International Linear 
Collider where hadronic photons will have large cross section in both $e^+e^-$ and 
$\gamma \gamma$ modes. Heavy quarks will be copiously produced at future colliders 
as a background to the more exotic processes expected. Therefore a precise description 
of their production properties within QCD will aid in the discovery of physics 
beyond the Standard Model. An example of this was studied by the ATLAS collaboration 
using Monte Carlo to simulate the production at the LHC of a $b\bar{b}$ pair along 
with a supersymmetric Higgs particle ($H/A$) which subsequently decays to a 
$b\bar{b}$ pair~\cite{atlas_tdr}. For an assumed mass $m_A =$~500\,GeV, even 
requiring four beauty jets, a signal-to-background ratio of only a few percent would 
be achieved. The irreducible background arises from QCD processes where the dominant 
processes are $gg$ and $gb$ with a gluon splitting to a $b\bar{b}$ pair. A discovery 
in this channel would therefore only be possible with precise knowledge these QCD 
background processes. 

%
%

\section{Theoretical and phenomenological overview}

For a generic collision between two hadrons, $H_a$ and $H_b$, in which a heavy quark 
pair is produced (see Fig.~\ref{fig:generic}),

\[ H_a + H_b \rightarrow Q \bar{Q}+ X, \]
the production cross section, $\sigma(S)$, for such a reaction at a centre-of-mass 
energy, $S$, can be written as:

\[ \sigma(S) = \sum_{i,j} \int dx_1 \int dx_2 \ \hat{\sigma}_{ij}(x_1 x_2 S, m^2, \mu^2)
                                            f_i^{H_a}(x_1, \mu) f_j^{H_b}(x_2, \mu), \]
where the right-hand side is a convolution of the parton densities in the colliding hadrons, 
$f_i^{H_a}$ and $f_j^{H_b}$, and the short-distance cross section, $\hat{\sigma}_{ij}$. These 
are evaluated at a renormalisation and factorisation scale, $\mu$, and momentum fractions 
of the colliding partons, $x_1$ and $x_2$. The parton densities are extracted from QCD fits 
to inclusive deep inelastic scattering and other data. The short-distance cross section 
is calculable in QCD and is a perturbative expansion in the mass of the heavy quark, $m$:

\[ \hat{\sigma}_{ij}(s, m^2, \mu^2) = \frac{\alpha_s^2(\mu^2)}{m^2} \left[ f_{ij}^{(0)}(\rho)
                                    + 4\pi\alpha_s(\mu^2)
           \left[f_{ij}^{(1)}(\rho) +\bar{f}_{ij}^{(1)}(\rho) \log(\mu^2/m^2) \right]
                                    + \mathcal{O}\mathit(\alpha_s^{\mathrm 2}) \right],
                                    \ \ \ \rho = 4m^2/s. \]
The expansion demonstrates that the larger the mass the faster the convergence. Hence 
predictions for beauty production should be more accurate than those for charm.

The treatment of the mass of the heavy quark is an important consideration for the 
implementation of the perturbative formalism in calculations. There are three schemes used:
the fixed-order (FO) or ``massive'' scheme, the resummed to next-to-leading logarithms (NLL),  
or ``massless'' scheme and more recently a scheme matching the two, known as 
FONLL~\cite{jhep:9805:007}. In the 
FO scheme, the predictions should be valid for transverse momenta of the order of 
the mass of the heavy quark. In this scheme, the heavy quarks are not active flavours 
in the parton distributions of the incoming hadron(s); they are produced in the hard 
scatter through processes such as $gg \to Q \bar{Q}$ shown in 
Fig.~\ref{fig:generic}. The resummed scheme is valid for transverse momenta 
much larger than the heavy quark mass. The heavy quarks are active flavours in the 
parton distributions of the incoming hadron(s), so can be produced by reactions such as 
$gQ \to gQ$. The FONLL calculations match the two schemes and are valid for 
all transverse momenta. The validity of the different calculations is investigated in 
comparison with data, particularly as a function the energy scale.

The fixed-order calculations used are from Frixione et al. 
(FMNR)~\cite{pl:b348:633,*np:b454:3} for photoproduction processes and {\sc Hvqdis} from 
Harris and Smith~\cite{pr:d57:2806} for deep inelastic scattering. Resummed calculations 
are only available for photoproduction at HERA from two groups of authors, Cacciari 
et al.~\cite{pr:d55:2736,*pr:d55:7134} and Kniehl et 
al.~\cite{zfp:c76:677,*zfp:c76:689,*pr:d58:014014,*pr:d70:094035}. The FONLL calculation 
is also only available in photoproduction. A calculation which is already available for 
some processes in $pp$ collisions, MC@NLO~\cite{jhep:0206:029,*jhep:0308:007}, combines 
a fixed-order calculation with the parton showering and hadronisation from the 
{\sc Herwig} Monte Carlo generator~\cite{jhep:0101:010}. Processes at HERA are not yet included, 
but it is hoped they will be done in the future and thereby provide a new level of detail in 
comparison with experimental data.

The advantages of a programme such as MC@NLO are its simulation of higher orders and also 
its sophisticated approach to hadronisation which attempts to describe the whole of the 
final state. The other programmes produce partons in the final state and fragment the outgoing 
quark to a hadron usually via the Peterson function~\cite{pr:d27:105}. Therefore these 
calculations may not be able to describe the full hadronic final state of an event. The validity 
of the fragmentation functions used also needs to be tested; they are usually extracted from 
fits to $e^+e^-$ data and their applicability to $ep$ or $pp$ needs to be demonstrated. 
Therefore the fragmentation function should be measured at HERA, and is discussed later,  
or measurements need to be made at high transverse energy or using jets where the effects of 
fragmentation are reduced.

Hadron-hadron collisions producing heavy quarks pairs can be simplified to and provide 
information on: the parton densities and in particular the gluon and heavy quark content 
of the hadron; the hard scatter and the dynamics of QCD as implemented into programmes; 
fragmentation or description of the parton to hadron transition. All of these aspects are 
discussed in this write-up.

%
%

\section{Information needed by the LHC experiments}

The information needed by the LHC which can be provided by the HERA experiments is the 
following:

\begin{itemize}

\item the state of the description of heavy quark production data by theoretical predictions. 
      The production of heavy quarks in the hard scattering process is discussed here in 
      detail. Information on heavy quarks produced in the splitting of a gluon outgoing from 
      the hard sub-process is also important for the LHC, but the information from HERA is 
      currently limited;

\item the gluon and heavy quark content of the proton parton density functions;

\item details of fragmentation in a hadronic environment;

\item the effect of the underlying event in heavy quark processes. This information is 
      limited at HERA but may be studied in the future;

\item HERA results can provide general information on event and jet topologies which will 
      be useful for designing algorithms or triggers at the LHC experiments.

\end{itemize}

The designing of effective triggers for $b$ physics is particularly acute for the LHCb 
experiment~\cite{misc:brook:private}. Large backgrounds are expected although event topologies 
should be 
different to the signal $b$ physics. For example minimum bias events will have a smaller 
track multiplicity and a lower transverse momentum for the highest $p_T$ track. Therefore 
using Monte Carlo simulation, cuts can be found to be able to reduce the rate of minimum bias 
whilst triggering efficiently on $b$ events. Such simulations require reliable Monte Carlo 
simulation of the event topologies of both classes of events.

Measurements of the proton structure function at HERA will constrain the parton densities 
in a large region of the kinematic plane where $B$ mesons will be produced within the 
acceptance of the LHCb detector. According to Monte Carlo simulations, these 
events are produced predominantly with a $b$ quark in the proton. However, this is just a 
model ({\sc Pythia}~\cite{cpc:135:238}) and at NLO some of the events will be summed into the gluon 
distribution of the proton. Nevertheless, measuring all flavours in the proton at HERA 
is one of the goals of the experiments and recent results on the beauty contribution 
to the proton structure function~\cite{epj:c40:349,*desy-05-110} shed some light on the issue. 

%
%

\section{Open beauty production}

\begin{figure}[htp]
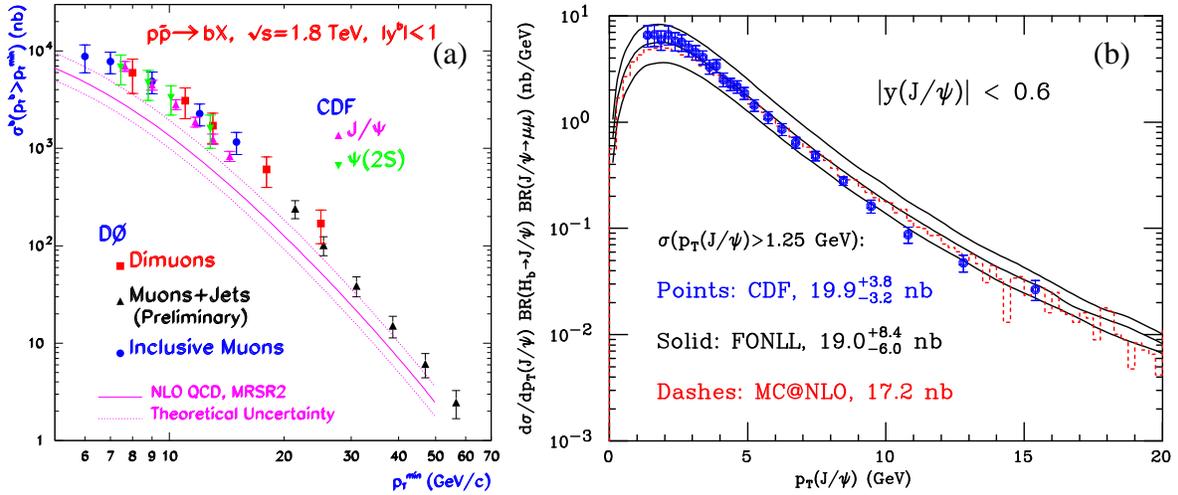

\begin{center}
~\epsfig{file=Figures/CDF-D0_b_Xsect.epsi,height=6.5cm}
~\epsfig{file=Figures/cdfpsi.eps,height=6.5cm}
\put(-280,170){\makebox(0,0)[tl]{\large (a)}}
\put(-30,170){\makebox(0,0)[tl]{\large (b)}}
\end{center}
\caption{(a) Tevatron Run I data extrapolated and compared to NLO QCD predictions and (b) 
         Run II data presented in terms of the measured quantities and compared to improved QCD 
	 theory.}
\label{fig:b_tev}
\end{figure}

The production of open beauty and its description by QCD has been of great interest in the 
last 10--15 years.The difference between the rates observed by the Tevatron 
experiments~\cite{prl:71:500,*prl:71:2396,*prl:75:1451,*pr:d53:1051,*pr:d65:052005,*prl:74:3548,*pl:b487:264,*prl:84:5478,*prl:85:5068} and NLO QCD predictions led to a mini crisis with many 
explanations put forward. Several measurements were performed in different decay channels and 
then extrapolated to the quark level to facilitate a comparison with QCD and between themselves. 
The NLO QCD prediction was found to be a factor of 2--3 below the data for all measurements 
as shown in Fig.~\ref{fig:b_tev}a. As mentioned, these results were extrapolated to the $b$-quark 
level using Monte Carlo models which may or may not give a good estimate of this extrapolation. 
To facilitate a particular comparison, an extrapolation can be useful, but should always be treated 
with caution and the procedure clearly stated and values of extrapolation factors given. Initial 
measurements in terms of measured quantities should also always be given.

The CDF collaboration also published measurements of $B$ meson cross sections. They were also 
found to be significantly above NLO calculations, but allowed for phenomenological study. Work 
on the fragmentation function was performed by Cacciari and Nason~\cite{prl:89:122003} which 
in combination with updated parton density functions and the FONLL calculation gave an increased 
prediction. New measurements at Run II have also been made by the CDF collaboration which probe 
down to very low transverse momenta. In combination with a measured cross section lower (but 
consistent) than the Run I data, and the above theoretical improvements, the data and theory 
now agree very well 
as shown in Fig.~\ref{fig:b_tev}b. The programme MC@NLO also gives a good description of the 
data.

\begin{figure}[htp]
\begin{center}
~\epsfig{file=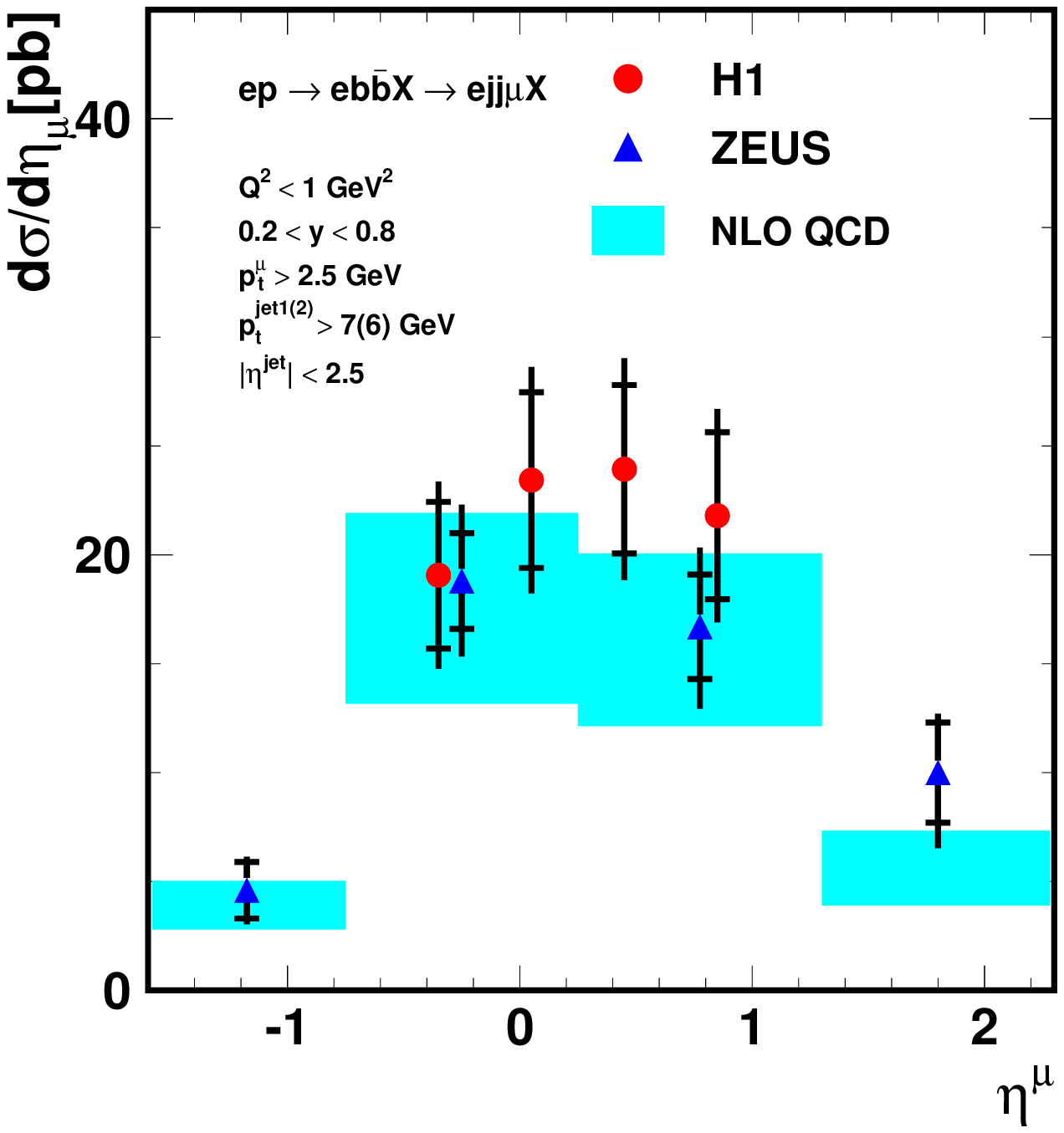,height=8cm}
~\epsfig{file=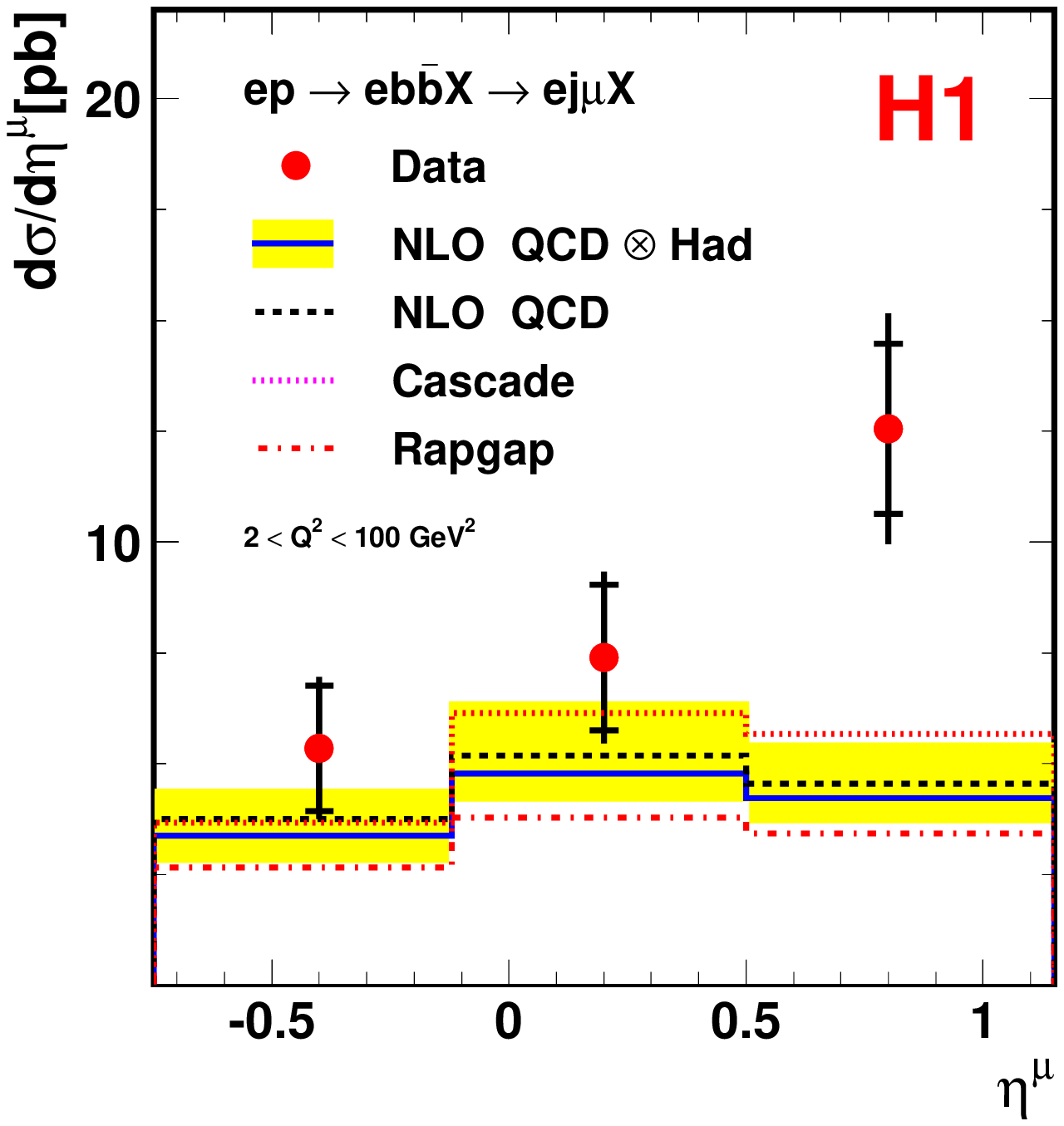,height=8cm}
\put(-260,190){\makebox(0,0)[tl]{\large (a)}}
\put(-30,190){\makebox(0,0)[tl]{\large (b)}}
\end{center}
\caption{Measurement of open beauty production as a function of the pseudorapidity of the decay 
         muon for (a) dijet photoproduction from the H1 and ZEUS experiments and (b) inclusive 
	 jet deep inelastic scattering from the H1 experiment. (The measurement from ZEUS 
	 experiment for (b) is in a different kinematic region but reveals the same physics 
	 message and so for brevity, is not shown)}
\label{fig:b_hera}
\end{figure}

The first result from HERA~\cite{pl:b467:156} also revealed a large discrepancy with NLO 
QCD predictions. This analysis also presented an extrapolated quantity, whereas later 
measurements~\cite{epj:c18:625,pr:d70:012008,*epj:c41:453,pl:b599:173} also presented measured 
quantities. The most recent and precise measurements~\cite{pr:d70:012008,*epj:c41:453} of beauty 
production with accompanying jets are shown 
in Fig.~\ref{fig:b_hera} compared with predictions from NLO QCD. The measurements in 
photoproduction (Fig.~\ref{fig:b_hera}a) are shown to be very well described by the prediction 
and the data from the two collaborations also agree well. The H1 data is somewhat higher than 
that from ZEUS; the difference is concentrated at low $p_T^\mu$ where the H1 data is also above 
the NLO calculation. The measurements in deep inelastic scattering are also generally described 
by NLO QCD although some differences at forward $\eta^\mu$ and low $p_T^\mu$ are observed by 
both collaborations. However, inclusive measurements which lead to a measurement of the beauty 
contribution to the proton structure function~\cite{epj:c40:349,*desy-05-110} are well described 
by QCD (see next Section).

The situation for the QCD description of $b$ production has recently changed significantly. In 
general, QCD provides a good description of the data with some hints at differences in specific 
regions. Certainly, there is no longer a difference of a factor of 2--3 independent of $p_T$. 
The HERA experiments will produce several new measurements in the next few years of higher 
precision and covering a larger kinematic region at both low and high $p_T$ and forward $\eta$. 
Allied with expected calculational and phenomenological improvements, a deep understanding of 
beauty production should be achieved by the turn-on of the LHC.

%
%

\section{Open charm production}

\begin{figure}[htp]
\begin{center}
~\epsfig{file=Figures/h1_zeus_eta.epsi,height=7cm}
~\epsfig{file=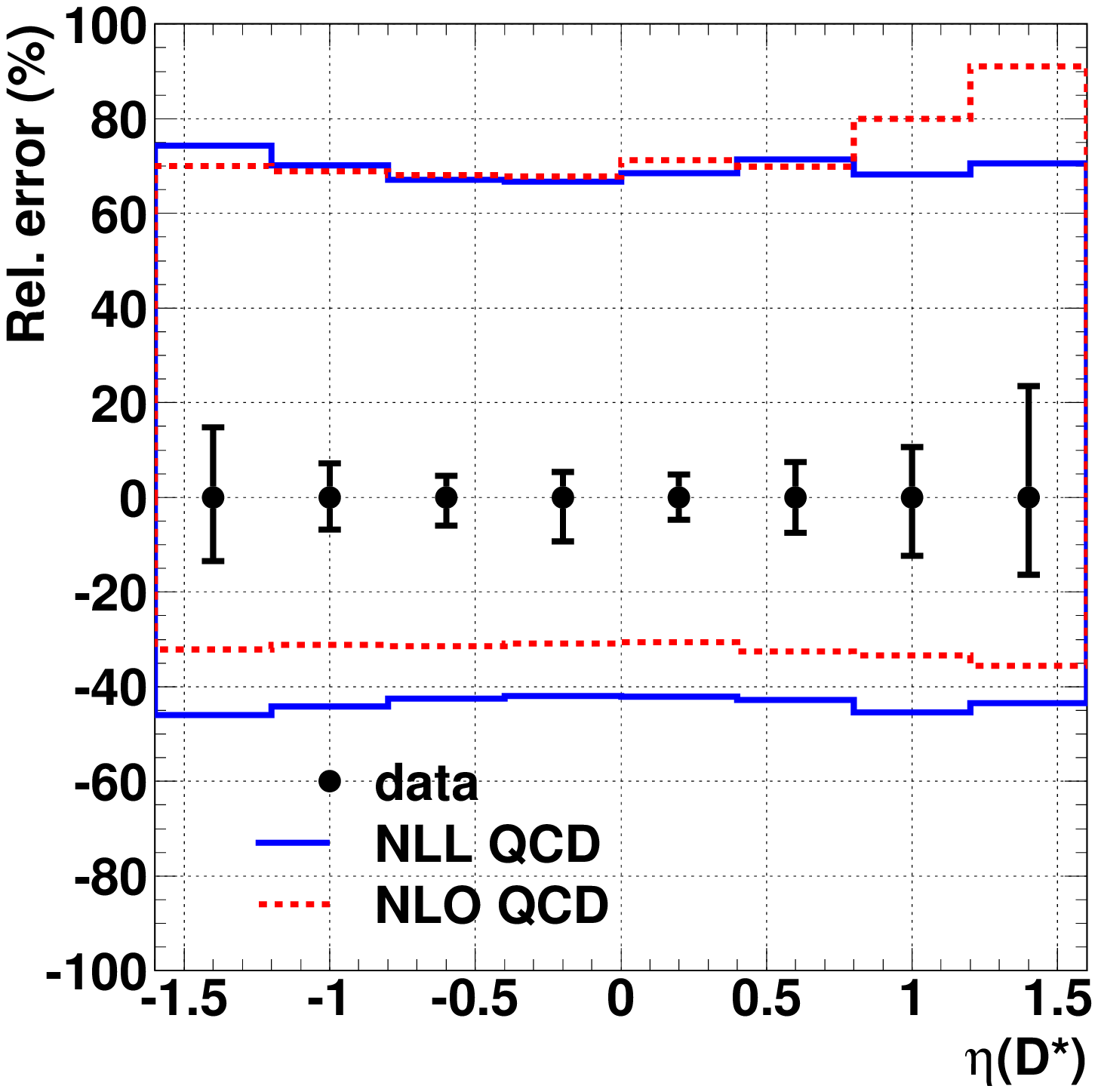,height=7cm}
\put(-230,160){\makebox(0,0)[tl]{\large (a)}}
\put(-30,160){\makebox(0,0)[tl]{\large (b)}}
\end{center}
\caption{Measurement of $D^*$ production compared with NLO QCD predictions: (a) 
         the differential cross section in deep inelastic scattering and (b) the relative 
	 uncertainty in data and theory in photoproduction.}
\label{fig:c_xsec}
\end{figure}

Due to its smaller mass, predictions for charm production are less accurate than for beauty. 
However large data samples allow detailed comparisons with theory. 
An example of a measured $D^*$ cross section in deep inelastic scattering is shown in 
Fig.~\ref{fig:c_xsec}a; data from the 
two experiments agree with each other and are well described by the prediction of QCD. Similar 
measurements have been made in photoproduction in which the data is less well described. Due to 
the larger cross section, the photoproduction data could prove valuable in constraining the photon 
as well as the proton structure. However, as can be seen from Fig.~\ref{fig:c_xsec}b, the 
theoretical precision is lagging well behind that of the data. Therefore more exclusive 
quantities and regions, with smaller theoretical uncertainties, are measured.

\begin{figure}[htp]
\begin{center}
~\epsfig{file=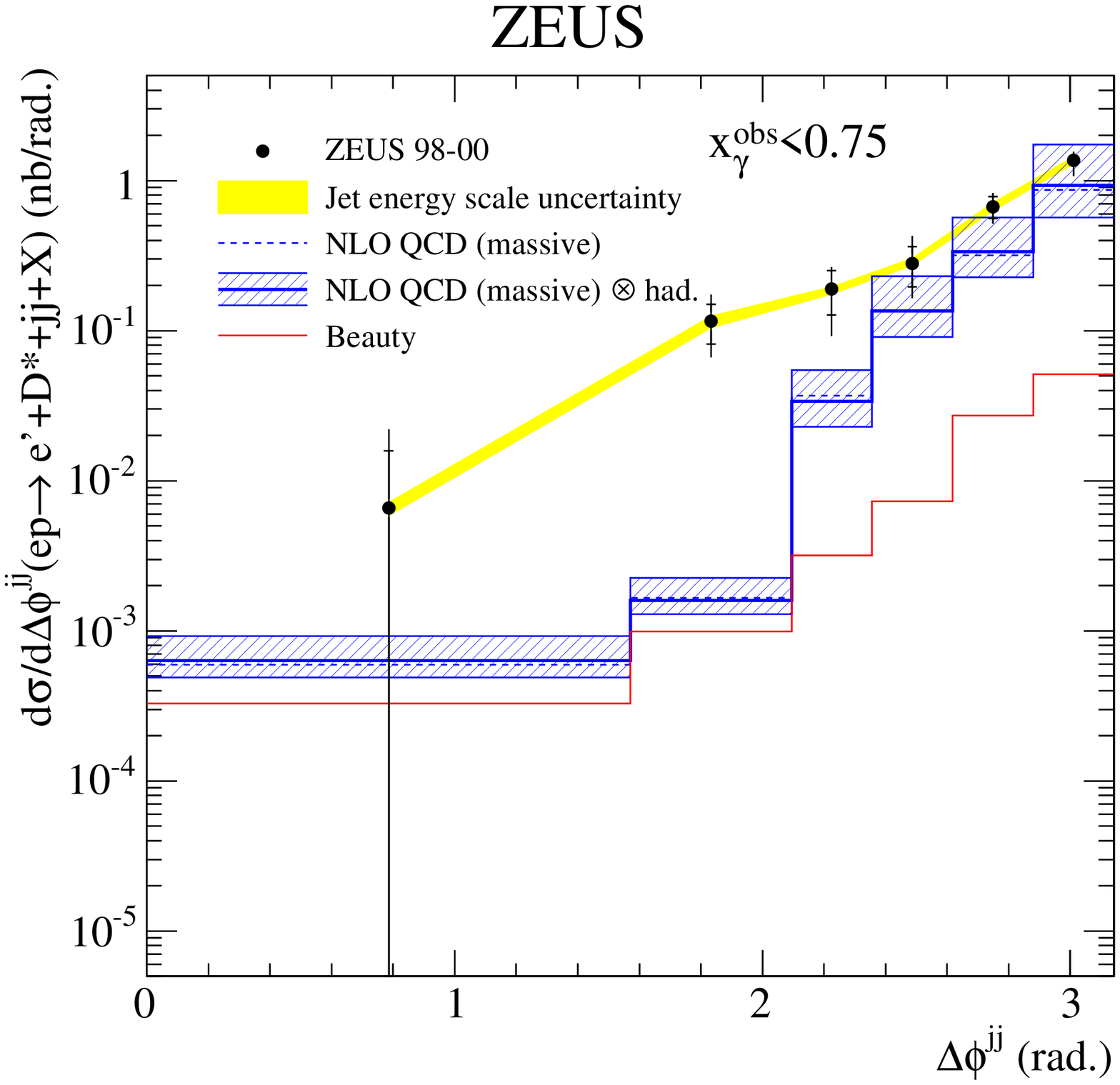,height=7cm}
~\epsfig{file=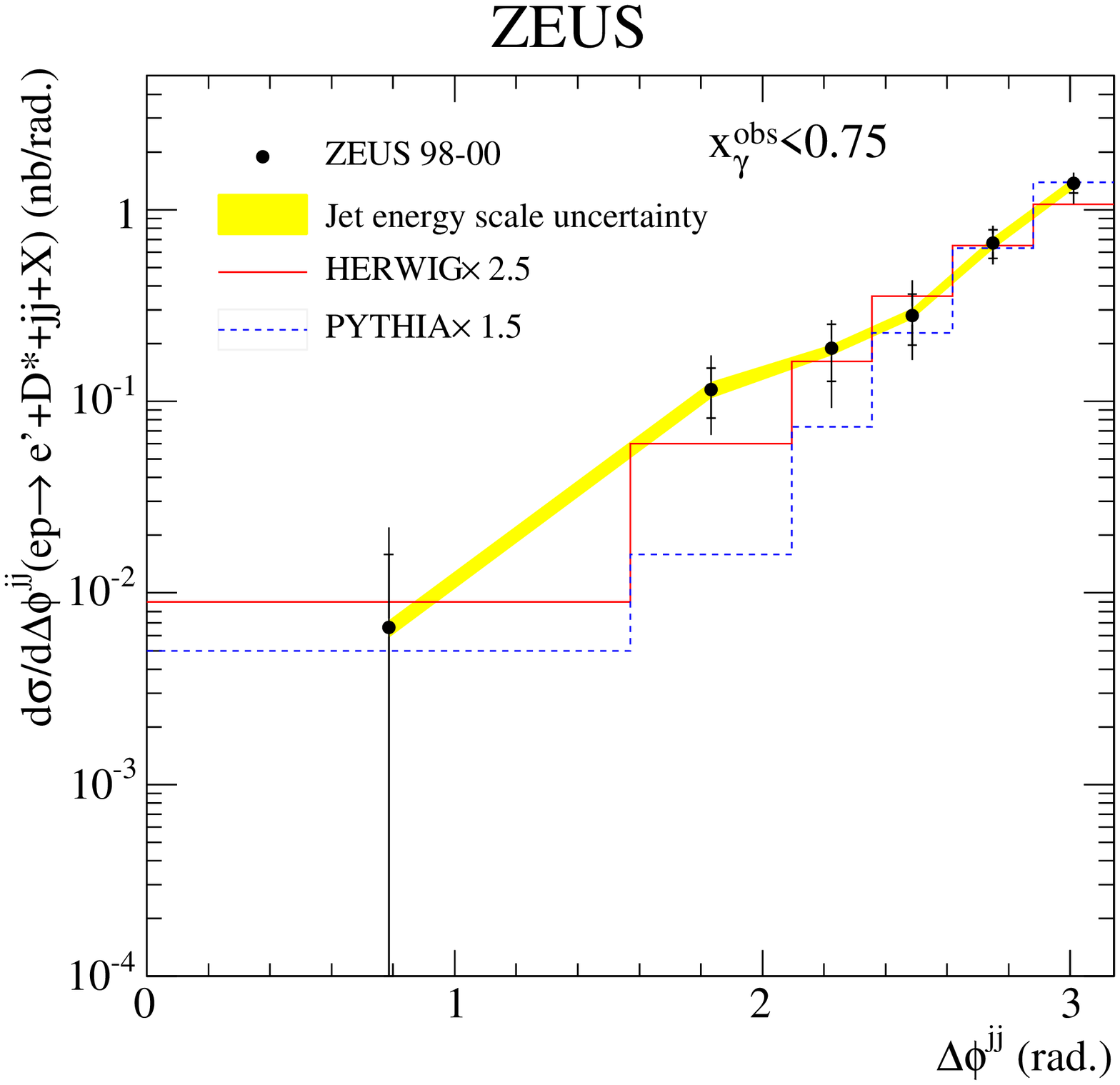,height=7cm}
\put(-245,40){\makebox(0,0)[tl]{\large (a)}}
\put(-30,40){\makebox(0,0)[tl]{\large (b)}}
\end{center}
\caption{Difference in the azimuthal angle of the two highest $E_T$ jets in charm 
         photoproduction for a sample enriched in resolved photon events compared to (a) a 
	 NLO QCD calculation  and (b) Monte Carlo models.}
\label{fig:c_jets}
\end{figure}

Measurements of charm photoproduction accompanied with jets pose a challenge for theory due 
the extra scale of the jet transverse energy. Such complicated final states will be copious 
at the LHC, so the verification of theory to HERA data will aid in the understanding of these 
high-rate QCD events. Dijet correlations in photoproduction have recently been 
measured~\cite{desy-05-132} and 
compared with available calculations. Events were selected in two regions: one enriched 
in direct photon events where the photon acts as a pointlike object and one enriched in 
resolved photon events where the photon acts as a source of partons. The cross section of 
the difference in the azimuthal angle, $\Delta \phi^{\rm jj}$, of the two highest $E_T$ 
jets has been measured. For the LO $2 \to 2$ process, the two jets are back-to-back. The 
data exhibit a significant cross section at low $\Delta \phi^{\rm jj}$ and for the 
direct photon events are reasonably well described by NLO QCD (not shown). However, the 
description for resolved photon events is poor as shown in Fig.~\ref{fig:c_jets}a. This 
region is particularly sensitive to higher orders not present in the QCD calculation. Monte 
Carlo models are compared to the data in Fig.~\ref{fig:c_jets}b; although the normalisation 
is poor, the shape of the distribution is very well described by the {\sc Herwig} simulation. 
This indicates that for the precise description of such processes, higher-order calculations 
or the implementation of additional parton showers in current NLO calculations are needed.

%
%

\section{The structure of the proton}

Open charm (and more recently beauty) production in deep inelastic scattering acts as a powerful 
probe of the structure of the proton, particularly the gluon and heavy quark densities. Such 
a direct measurement of the gluon density complements its extraction in QCD fits to inclusive 
data. The cross section for the production of a heavy quark pair can be written in terms of 
the heavy quark contribution to the proton structure functions:

\[
\frac{d^2\sigma^{Q\bar{Q}} \left(x, Q^2\right)}{dxdQ^2} =
\frac{2\pi\alpha^2}{x Q^4}
\left\{ \left[1+\left(1-y\right)^2 \right] F_2^{Q\bar{Q}}\left(x, Q^2\right)
- y^2 F_L^{Q\bar{Q}}\left(x, Q^2\right) \right\}
\]
The value of the charm contribution, $F_2^{c\bar{c}}$, has traditionally been extracted by 
measuring $D^*$ mesons within the acceptance of the detector and extrapolating to the full phase 
space. 

\begin{figure}[htp]
\begin{center}
~\epsfig{file=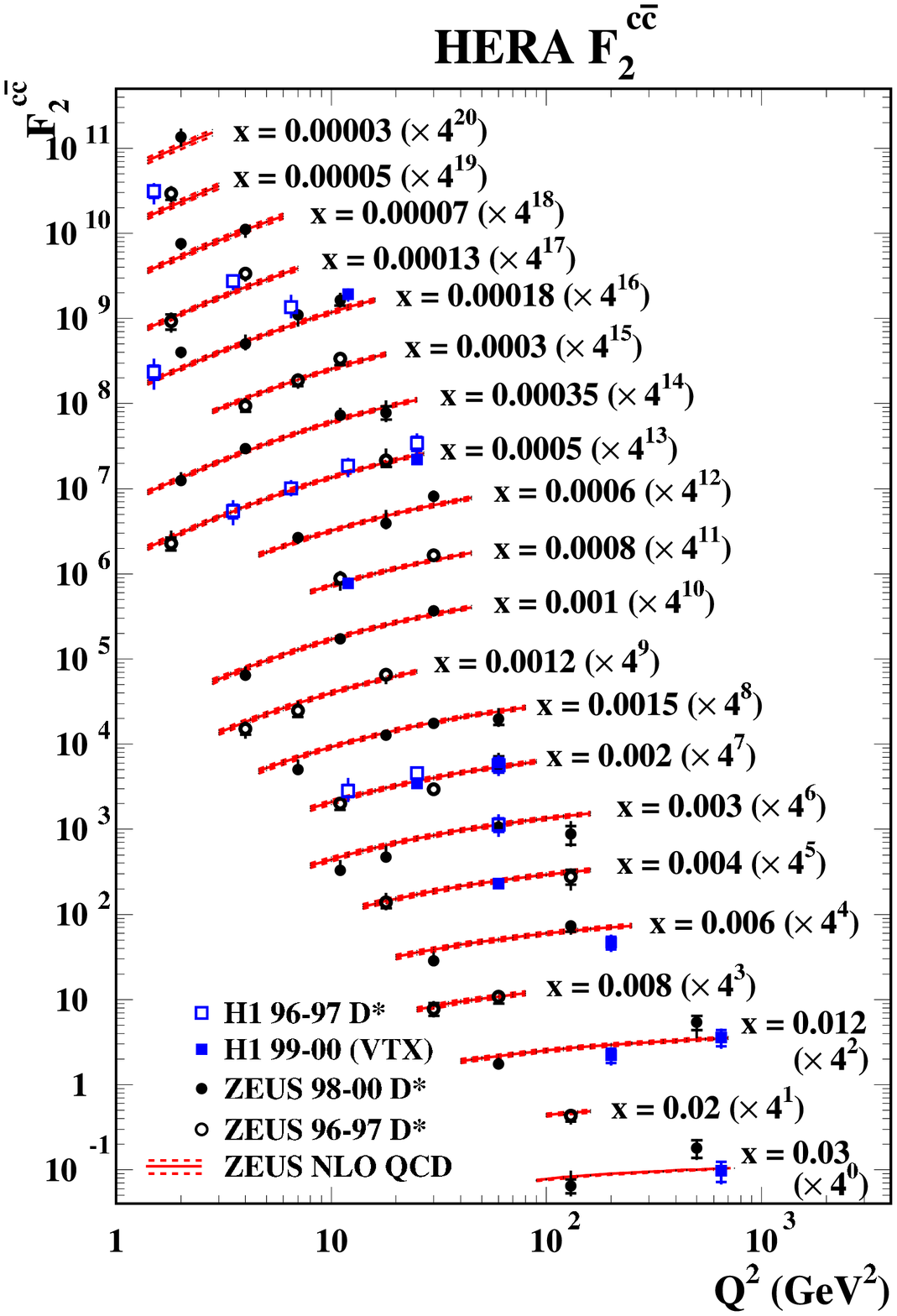,height=11cm}
~\epsfig{file=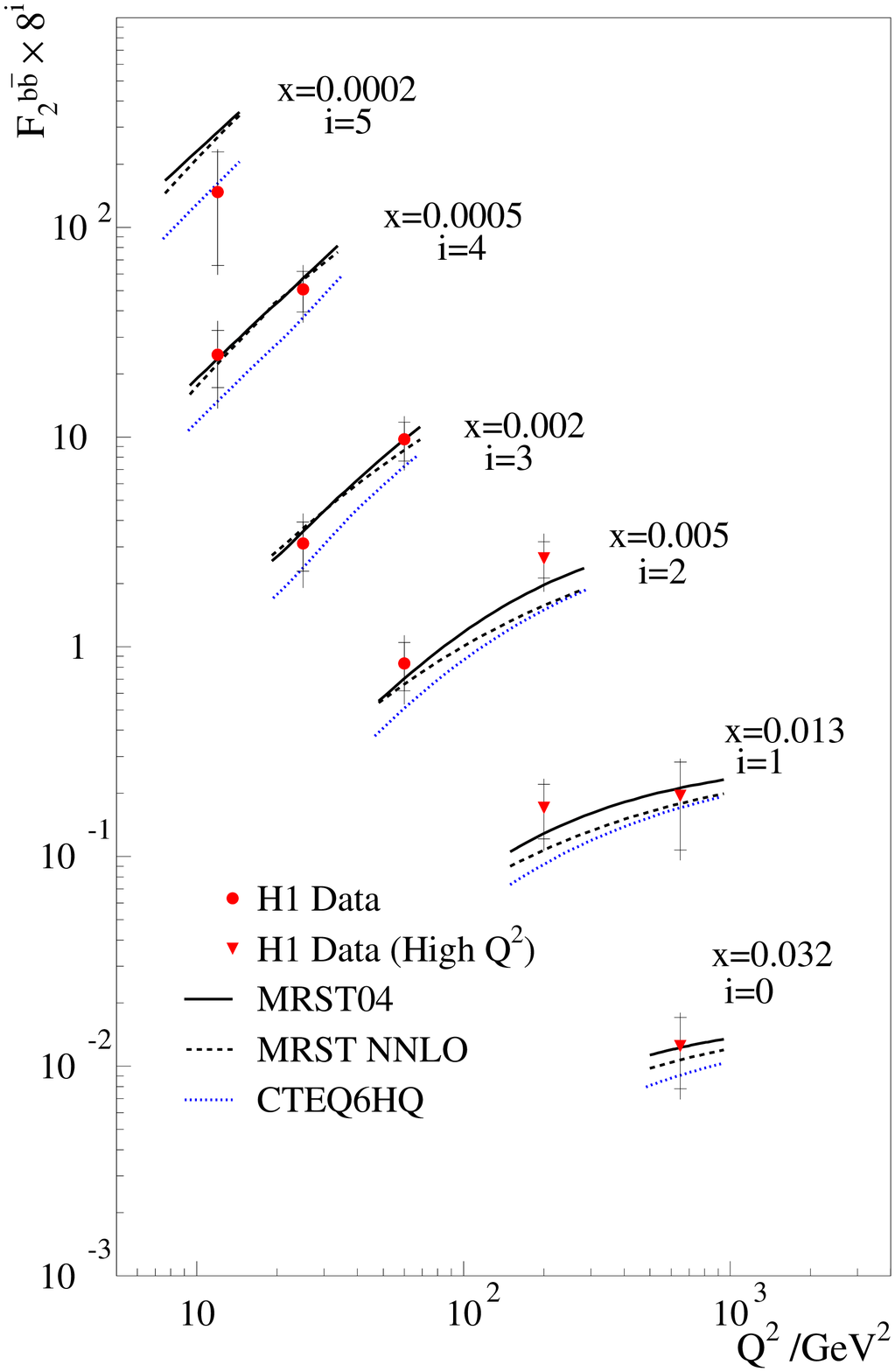,height=11cm}
\put(-240,280){\makebox(0,0)[tl]{\large (a)}}
\put(-30,280){\makebox(0,0)[tl]{\large (b)}}
\end{center}
\caption{(a) Charm contribution, $F_2^{c\bar{c}}$, and (b) beauty contribution, $F_2^{b\bar{b}}$, 
          to the proton structure function, $F_2$, versus $Q^2$ for fixed $x$.}
\label{fig:f2_qq}
\end{figure}

The values of $F_2^{c\bar{c}}$ extracted from the measured $D^*$ cross 
sections~\cite{epj:c12:35,pl:b528:199,pr:d69:012004} are shown in Fig.~\ref{fig:f2_qq}a compared 
with NLO QCD. New measurements of $F_2^{c\bar{c}}$ have been recently performed using an inclusive 
sample of high $p_T$ tracks~\cite{epj:c40:349,*desy-05-110}. This data is more inclusive than the 
$D^*$ measurements probing much lower $p_T$ and thereby having much reduced extrapolation factors 
(a factor of ~1.2 rather than 2--3 as for the $D^*$ measurements). These results confirm the 
previous data and add extra information on $F_2^{c\bar{c}}$. The results on $F_2^{c\bar{c}}$ 
demonstrate a large gluon density in the proton as exhibited by the scaling violations versus 
$Q^2$ and are well described by such a parton density function. At high $Q^2$, charm contributes 
up to about 30\% of the inclusive cross section. It is hoped with higher statistics and a better 
control over the systematics that the charm cross section data can be used in QCD fits to constrain 
the gluon (or heavy quark) density in the proton.

Applying the same technique of using high $p_T$ tracks, the H1 collaboration have made measurements 
of $F_2^{b\bar{b}}$ which are shown in Fig.~\ref{fig:f2_qq}. The results are consistent with 
scaling violations and are well described by new parton density functions. The differences between 
the different parametrisations are not insignificant and future measurements should be able to 
discriminate between them. For the $Q^2$ range measured, beauty production contributes up to 
3\% of the inclusive cross section.

%
%

\section{Universality of charm fragmentation}

\begin{figure}[htp]
\hspace{-0.45cm}
\begin{minipage}[b]{6.5cm}
~\epsfig{file=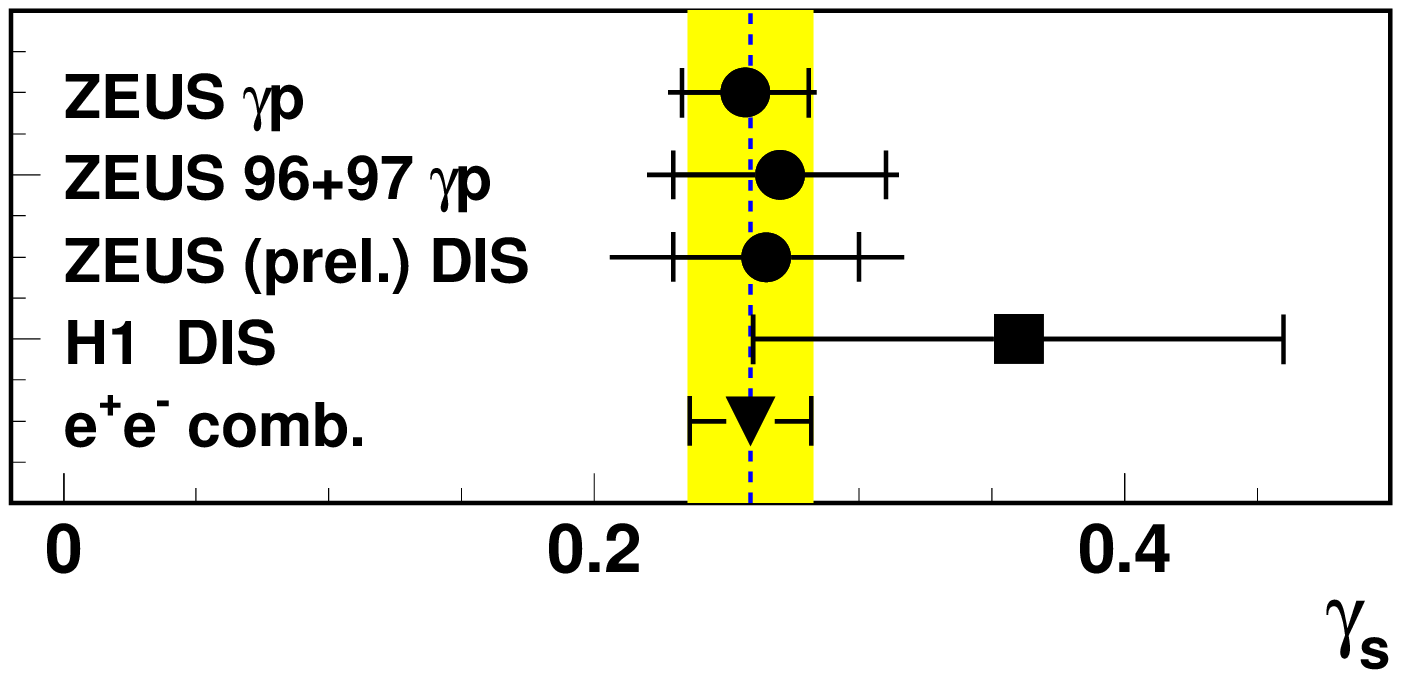,height=3.3cm}
~\epsfig{file=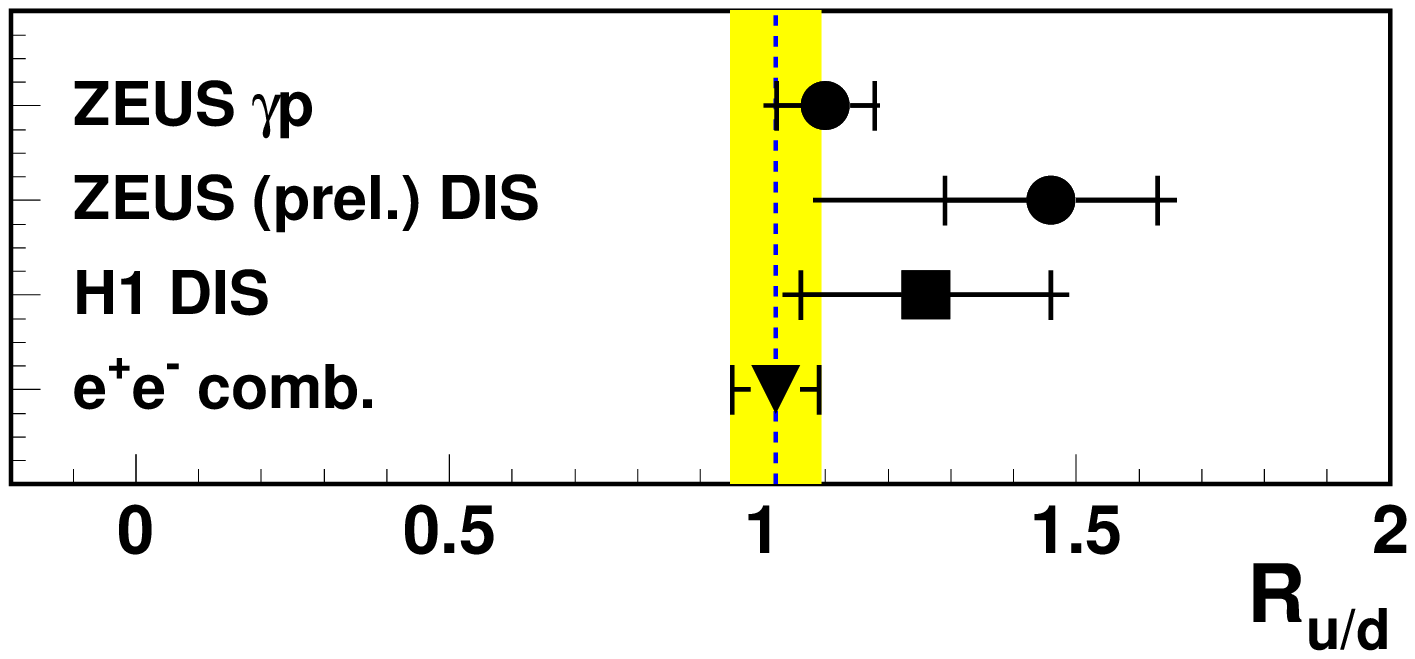,height=3.3cm}
~\epsfig{file=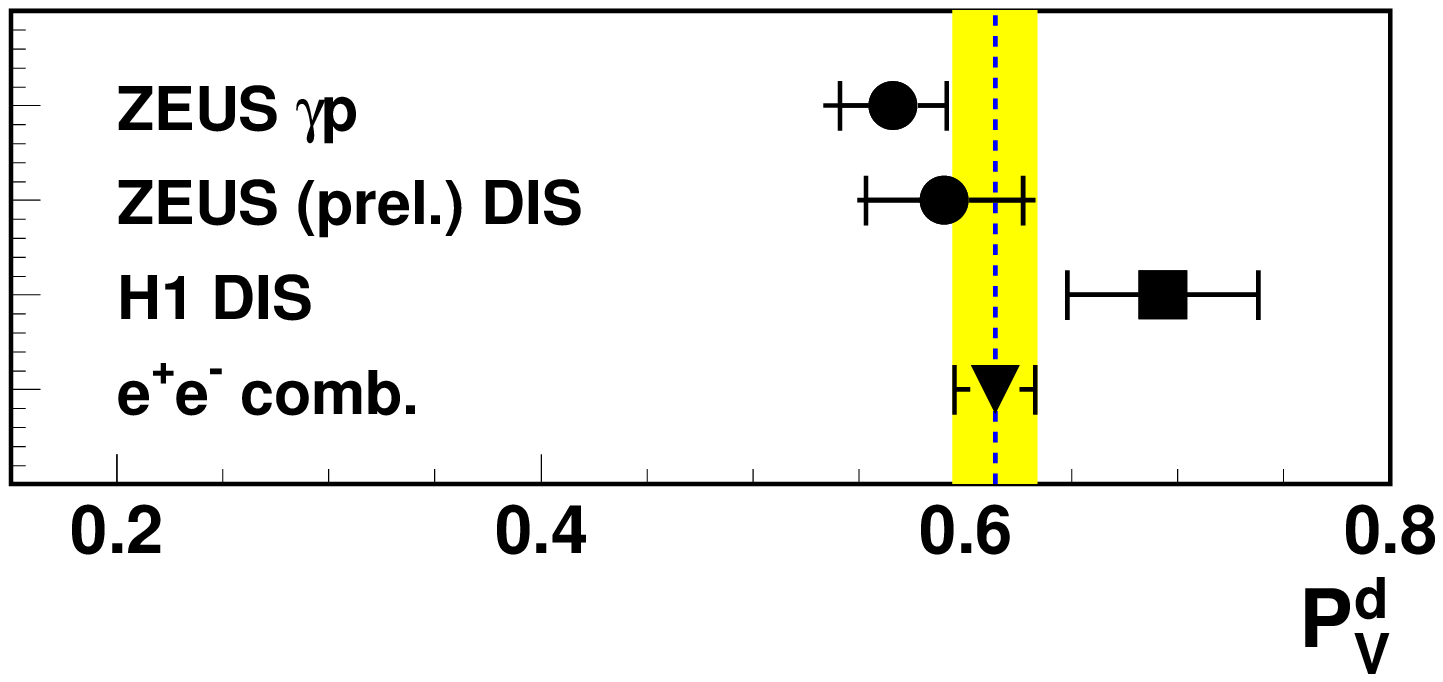,height=3.3cm}
\end{minipage}
\begin{minipage}[b]{8cm}
~\epsfig{file=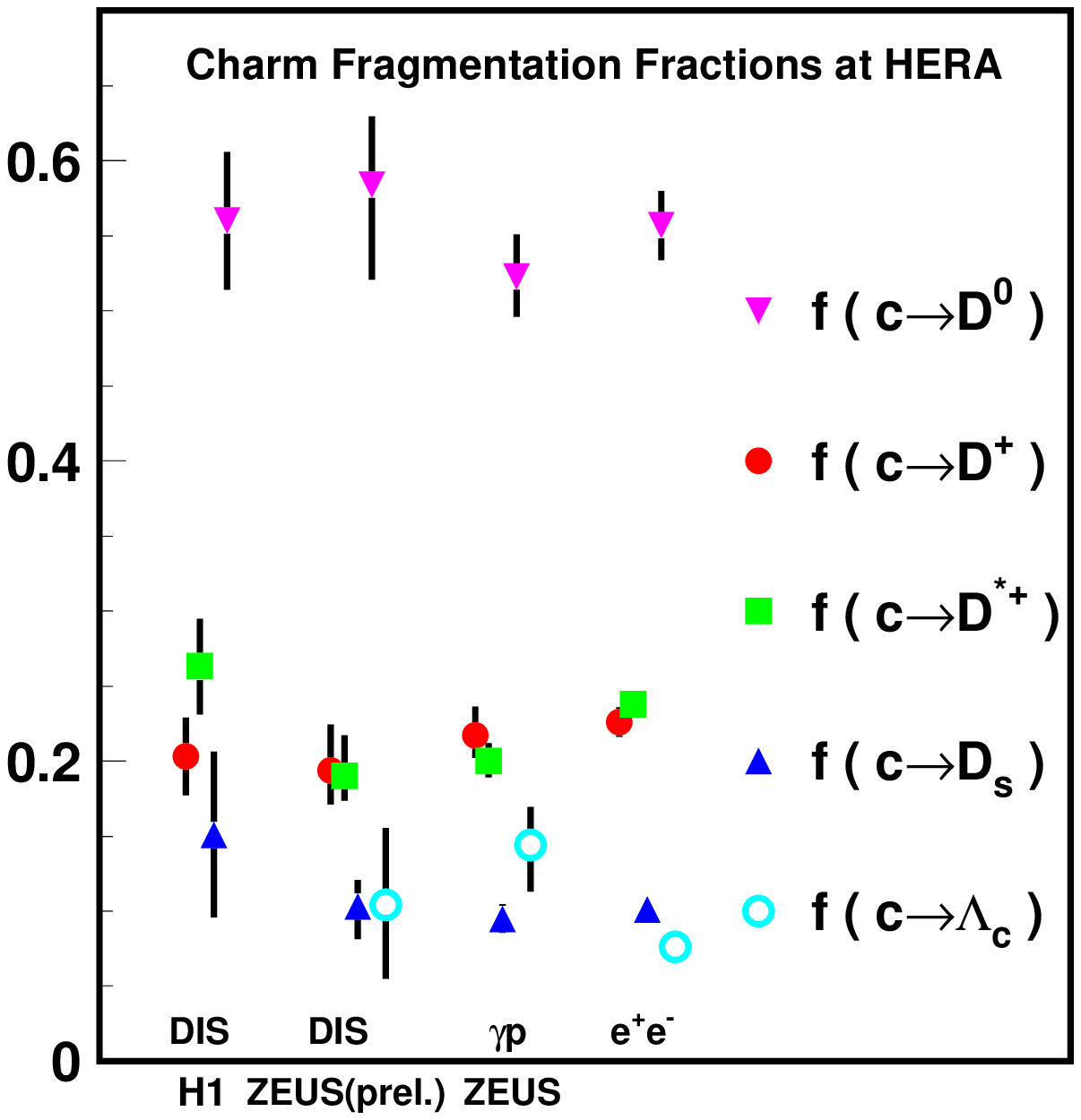,height=10cm}
\end{minipage}
\caption{Comparisons of fragmentation parameters, $\gamma_s$, $R_{u/d}$ and $P_v^d$
         and $f(c \to D, \Lambda)$ in photoproduction, deep inelastic scattering and $e^+e^-$ 
         collisions.}
\label{fig:frag}
\end{figure}

Heavy quark fragmentation has been extensively studied in $e^+e^-$ collisions. The clean 
environment, control over the centre-of-mass energy and back-to-back dijet system provide 
an ideal laboratory for accurate measurement of fragmentation parameters. The measured 
parameters, e.g. fragmentation function and fraction of charm quarks hadronising to a 
particular meson, are used as inputs to models and NLO QCD calculations of $ep$ collisions. 
Therefore, the validity of using fragmentation parameters extracted from $e^+e^-$ data 
in $ep$ data needs to be verified. The strangeness suppression factor, $\gamma_s$, the 
ratio of neutral and charged $D$-meson production rates, $R_{u/d}$, the fraction of charged 
$D$ mesons produced in a vector state, $P_v^d$ and the fragmentation fractions, 
$f(c \to D, \Lambda)$, have been measured in deep inelastic 
scattering~\cite{epj:c38:447,*cpaper:lp2005:266} and in photoproduction~\cite{desy-05-147}. 
The results are shown in Fig.~\ref{fig:frag} compared with values obtained in $e^+e^-$ 
collisions. The data obtained in different processes are consistent with each other 
and thereby consistent with the concept of universal fragmentation. The measurements in 
photoproduction also have precision competitive with the combined $e^+e^-$ data. The data 
therefore provide extra constraints and demonstrate that the fragmentation at a hadron 
collider in the central part of the detector looks like that in an $e^+e^-$ collision. This 
will provide useful input for future models to be used at the LHC.

The charm fragmentation function to $D^*$ mesons has been measured by both the H1 and ZEUS 
collaborations~\cite{cpaper:ichep2002:778,*cpaper:lp2005:407} and compared to $e^+e^-$ data. 
Although the definitions of the fragmentation 
function and the energies are different, the general trends are the same. However, a consistent 
fit to all data within a given Monte Carlo or NLO calculation is needed to clarify this 
situation. Measurements at the Tevatron would also contribute significantly to this area.

%
%

\section{Conclusions}

An increasing number of high precision measurements of heavy quark production from 
HERA have recently become available. They are providing valuable information on the parton 
densities, the overall production rates and the concept of the universality of fragmentation. 
Precise and well-defined measurements have allowed phenomenological improvements to be made. 
Generally QCD describes the production of heavy quarks; in particular, due in part to the 
advances made in the HERA measurements, the prediction for the production of beauty quarks is no 
longer well below the data. There are some details still lacking which await to be confronted with 
higher order calculations or NLO calculations interfaced with parton showers and hadronisation. 
There is also ongoing work in tuning Monte Carlo predictions to all known data which demonstrates 
the need to have global calculations which can predict all processes under study. In the next few  
years in the run up to the LHC, HERA will produce a lot more data and more will be known 
about heavy quark production.

\bibliographystyle{heralhc} 
{\raggedright
\bibliography{heralhc}
}
\end{document}